\documentclass[prb,twocolumn,showpacs,superscriptaddress,floatfix]{revtex4}
\usepackage{graphicx}
\usepackage{times,mathptmx}

\begin{document}

\title{Observation of Conduction Band Satellite of Ni Metal by 3$p$-3$d$ Resonant Inverse Photoemission Study}

\author{Y. Tezuka}
\affiliation{Faculty of Science and Technology, Hirosaki University, 3 Bunkyo-cho, Hirosaki, 036-8561, Japan}
\author{K. Kanai}
\author{H. Ishii}
\author{S. Shin}
\affiliation{Institute for Solid State Physics, University of Tokyo, 5-1-5 Kashiwanoha, Kashiwa-shi, Chiba 277-8581, Japan}
\author{S. Nozawa}
\affiliation{Department of Applied Physics, Tokyo University of Science, Tokyo 162-8601, Japan}
\author{A. Tanaka}
\author{T. Jo}
\affiliation{Department of Quantum Matter, ADSM, Hiroshima University, Higashi-Hiroshima 739-8526, Japan}

\date{\today}

\begin{abstract}
Resonant inverse photoemission spectra of Ni metal have been obtained across the Ni 3$p$ absorption edge. The intensity of Ni 3$d$ band just above Fermi edge shows asymmetric Fano-like resonance. Satellite structures are found at about 2.5 and 4.2 eV above Fermi edge, which show resonant enhancement at the absorption edge. The satellite structures are due to a many-body configuration interaction and confirms the existence of 3$d^8$ configuration in the ground state of Ni metal.
\end{abstract}

\maketitle

     Inverse photoemission spectroscopy (IPES) is an important technique to investigate the unoccupied density of states (DOS) of a solid. Combining photoemission spectroscopy (PES), which measures the occupied DOS, with IPES measurements, gives us complementary information about the valence and conduction band DOS.\cite{r01} The IPES technique has two measurement modes: Bremsstrahlung Isochromat Spectroscopy (BIS) mode and Tunable Photon Energy (TPE) mode. The BIS measurements are easier than TPE measurements, because it does not use a photon monochromator and sensitive band pass filters are available in X-ray and vacuum ultraviolet (VUV) region. This has led to the early development of X-ray BIS (XBIS) and ultra-violet BIS (UVBIS) techniques.\cite{r02,r03}

     The observation of IPES in the soft X-ray (SX) region corresponding to energies from several ten's of eV to about 1 keV is still experimentally difficult, because the emission intensity in IPE is extremely weak. We succeeded in the observation of the resonant IPES (RIPES) of Ce \cite{r04,r05} compounds near the Ce 4$d$ absorption region, using a monochromator developed for SX emission spectroscopy (SXES).\cite{r06} The obtained results are consistent with Ce 3$d$ RIPES by Weibel et al.\cite{r07}, though the surface effect is strong. Furthermore, RIPES of Ti \cite{r08,r09} compounds was also measured across the Ti 3$p$ edge and a weak satellite has been found.

     Ni metal is an itinerant ferromagnet which has been used as a classic reference to test the validity of new experimental and theoretical techniques in the study of electronic structure of solids. Beginning with the Stoner condition in the mean-field-approximation or the local density approximation (LDA) \cite{r10}, as well as many spectroscopic studies of Ni metal have provided important insights in the study of solids e.g. resonant PES \cite{r11,r12,r13}, angle-resolved PES \cite{r14}, magnetic circular dichroism (MCD) \cite{r15,r16}, and spin-resolved PES.\cite{r17,r18,r19} Furthermore, UVBIS \cite{r20} and XBIS \cite{r21} spectra of Ni metal have also been reported, as well as spin polarized \cite{r22,r23,r24} and k-resolved \cite{r25,r26,r27} IPES. The observed electronic structure of Ni is, however, still an important subject of study that many researchers are interested in, since it is not understood within standard band theory and only recent dynamical mean field studies \cite{r28} provide a consistent description of its magnetic properties and electronic structure.

     It is well known that the so-called "6-eV satellite" is observed in the PES spectrum at about 6 eV from Fermi energy $E_F$.\cite{r11,r17} This satellite is known as the two-hole-bound state that means two 3$d$ holes are bound in the same Ni site in the final state, and has a 3$d^8$ final state (3$d^9$ initial state).\cite{r29} Another satellite was found at a higher energy than 6 eV and it was assigned to the 3$d^7$ final state (3$d^8$ initial state).\cite{r30} Furthermore, it was suggested by analysis of the MCD spectra that the 3$d^8$ configuration with $^3F$ symmetry exists with a weight of 15 $\sim$ 20 \% in the ground state.\cite{r31,r32} Sinkovic et al. found triplet feature of 3$d^8$ configuration at 6 eV by means of spin-resolved PES.\cite{r19}

     The main 3$d$ configuration of Ni atom in Ni metal is 3$d^9$ in the ground state. From a many-body view-point, 3$d^{10}$ and 3$d^8$ should be mixed in addition to 3$d^9$ due to the electron transfer. Then, the ferromagnetism is considered to be caused by Hund's coupling in the 3$d^8$ configuration as it reduces the energy cost of an electron transfer. In fact, such a viewpoint is proposed as an origin of ferromagnetism in Ni.\cite{r33} In this context, an experimental measurement of 3$d^8$ weight is of great importance.

     In this study, we report resonant IPES of Ni metal across the Ni 3$p$-absorption edge. Since the process of the IPES adds an electron to the ground state, IPES should give us new information of the ground state configuration.

\begin{figure}
\includegraphics[scale=0.40]{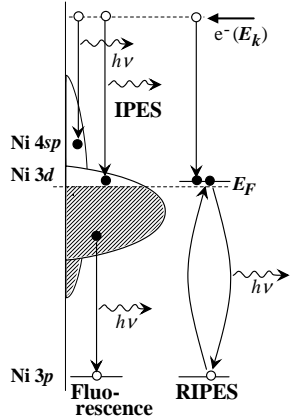}
\caption{\label{fig:epsart} Energy diagram of RIPES. At low energy excitation, only normal IPES is observed. If the excitation is higher than an absorption edge, a core hole is created which decays by fluorescence. Near an absorption edge, RIPES process can take place.}
\end{figure}

     Figure 1 shows energy diagram of RIPES. In a normal IPES process, an electron that is incident upon a solid surface decays radiatively to states at lower energy. In a 3$d^n$-electron system, the normal IPES process is expressed as
\begin{equation}
 |3d^n \rangle +e^- \longrightarrow |3d^{n+1} \rangle + h\nu
\end{equation}
where $e^-$ denotes incident electron. If the electron energy is higher than the binding energy of a core level, the core electron can be excited and ejected out of the system. Then, the created core-hole decays radiatively (fluorescence) or non-radiatively (Auger process). The fluorescence process is 
\begin{equation}
 |\underline{c}3d^n \rangle \longrightarrow |3d^{n-1} \rangle + h\nu
\end{equation}
where $\underline{c}$ denotes core hole. On the other hand, if the energy of the incident electron is close to the Ni 3$p$ $\rightarrow$ 3$d$ absorption edge, a second order process
\begin{equation}
 |3d^n \rangle +e^- \longrightarrow |3p^53d^{n+2} \rangle \longrightarrow |3d^{n+1} \rangle + h\nu
\end{equation}
would take place. Because of the interference between (1) and (3), a resonance effect would be observed.

     IPES measurements of Ni were performed on both polycrystal and (110) single crystal. The polycrystalline sample was evaporated on Mo substrate at a pressure of $< 1 \times 10^{-8}$ torr. Measurements were performed at low temperature of about 14 K. The cleanliness of the sample was checked by measuring O 1$s$ fluorescence. The measurement chamber pressure was $< 3 \times 10^{-10}$ torr throughout the measurements. Single crystal was measured with some excitation energies. (110) sample was cleaned by Ar-ion bombardment and annealing. The cleanliness was checked by Auger and LEED measurements.
 
     A soft X-ray monochromator, which consists of a Rowland-type grazing-incidence monochromator with a 5-m spherical grating (300 lines/mm), was used in this experiment.\cite{r05,r06} The incidence angle of monochromator was fixed at an angle of 85.98$^\circ$. A two types multi-channel detector PIAS (for wide range) and CR-chain (for high resolution) (Hamamatsu photonics) were used as a photon detector. The absolute energies of the spectra were calibrated by measuring the Fermi edge of Au.

     A filament-cathode-type and a BaO-cathode-type electron guns were used for excitation. The kinetic energy of excitation electron was calibrated by an energy analyzer. An excitation electron was incident normally for polycrystal, while off-normal for Ni(110), because of experimental arrangement. The emission was observed at an angle of about 60$^\circ$. The overall spectral resolution of this measurement was about 0.6 eV at excitation energy of 60 eV. The spectra were normalized by emission of electron gun and $(h\nu)^3$, since the cross section of emission spectra is proportional to third power of photon energy.

\begin{figure}
\includegraphics[scale=0.55]{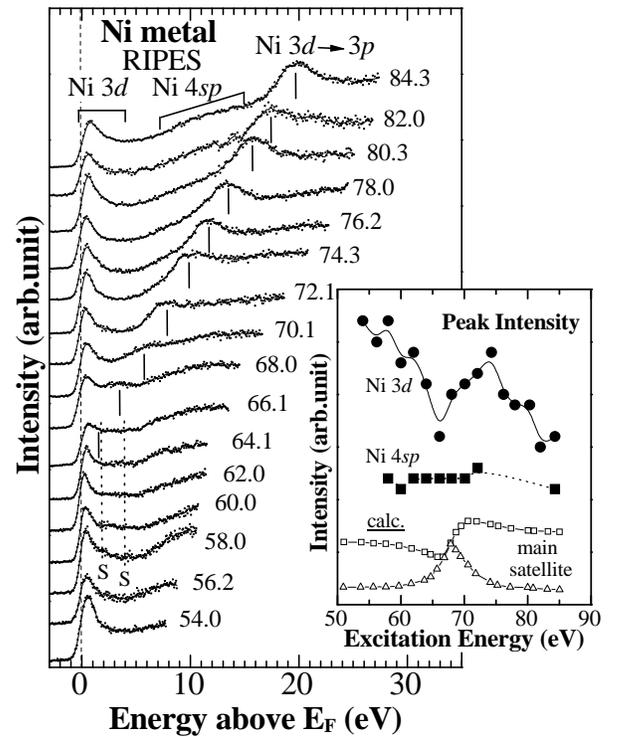}
\caption{\label{fig:epsart} RIPES spectra of Ni metal. The dots show observed spectra and solid lines were obtained by smoothing. The numbers beside the spectra denote the excitation energy. The energy positions of Ni 3$d$ $\rightarrow$ 3$p$ fluorescence in each spectrum are shown by vertical bars. Ni 3$d$ IPES peak is observed just above $E_F$. Ni 4$sp$ peak is also observed at about 10 eV. Dotted lines denote the satellite structures observed near the absorption edge. Insertion shows the intensity of IPES features. The filled circles and squares show the intensity of Ni 3$d$ main peak and Ni 4$sp$, respectively. The solid lines were obtained by smoothing, plotted as a guide for eyes. The open squares and triangles show calculated intensity of main peak and satellite, respectively.}
\end{figure}

     Figure 2 shows RIPES spectra of polycrystalline sample, obtained for various energies across the Ni 3$p$ absorption edge. Numbers beside the spectra indicate excitation energies. In this figure, observed spectra, which have energies close to excitation energies, are plotted with respect to the relative energy from Fermi edge. The spectrum of 54.0 eV, which is sufficiently below the absorption edge, corresponds to normal IPES spectrum. This spectrum agrees with the spectra observed in XBIS \cite{r20} and UVBIS. \cite{r21} From comparison with band calculations \cite{r34}, the structure just above Fermi edge and broad peak at about 10 eV are assigned to Ni 3$d$ and Ni 4$sp$ bands, respectively.

     When the excitation energy is higher than 66.1 eV, a core electron is excited. Thus, the emission spectrum then includes both IPES and fluorescence components. The Ni 3$d$ $\rightarrow$ 3$p$ fluorescence peak is observed at a constant energy of about 65 eV in emission spectra. The energy position of this peak is changed with changing excitation energy in Fig. 2 as indicated by vertical bars. The Ni 3$d$ peak just above $E_F$ becomes very weak when the excitation energy is around 66.1 eV, where the fluorescence peak has almost same emission energy. On the other hand, Ni 4$sp$ peak does not seem to change its intensity with changing excitation energy. In addition to these structures, a weak structure is observed at around 2.5 and 4.2 eV as indicated by the dotted line. These structures are observed only at the excitation near absorption edge.

     Insertion in Fig.2 shows the peak intensity of the Ni 3$d$ and Ni 4$sp$ peak plotted versus the excitation energy. Filled circles and squares denote the intensity of Ni 3$d$ and Ni 4$sp$, respectively. The open squares and triangles are calculated intensity that is discussed below.\cite{r35} The Ni 3$d$ spectrum has a dip at about 66 eV and shows an asymmetric lineshape typical of a Fano-type resonance.\cite{r36} A similar resonance has been observed in the resonant photoemission study of Ni.\cite{r12,r30} On the other hand, the Ni 4$sp$ peak does not change its intensity with changing excitation energy, although at higher energies it cannot be conclusively stated because of an overlap with the fluorescence signal. Thus, satellite intensity is observed at about 58 $\sim$ 66 eV spectrum, the peak intensity of these satellites is enhanced as seen in Fig. 2.

     The results show that the IPES of Ni 3$d$ exhibits a resonance effect at the excitation energy near Ni 3$p$-absorption edge. The nominal ground state of Ni is 3$d^9$ configuration. It is thought, however, that the actual ground state consists of a mixture of 3$d^8$, 3$d^9$ and 3$d^{10}$ configurations. The intermediate state of RIPES has an n+2 electron state as has been mentioned before. So, only the 3$d^8$ initial state can be resonant in the IPES process, while the 3$d^9$ and 3$d^{10}$ initial states cannot resonate. That is, the observed resonance confirms the existence of 3$d^8$ configuration in the ground state. The existence of 3$d^8$ configuration has been suggested by resonant PES \cite{r30} and MCD \cite{r31,r32} measurements. However, the present result is the only direct experimental evidence of a 3$d^8$ initial-state configuration.
 
\begin{figure}
\includegraphics[scale=0.55]{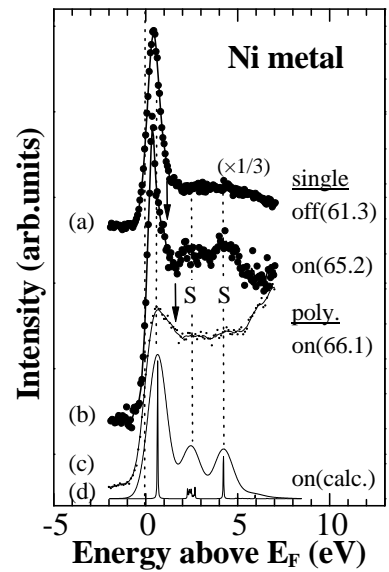}
\caption{\label{fig:epsart} Comparison of on- and off-resonant spectra. (a) off-resonant spectrum of single crystal, (b) on-resonant spectrum of single crystal, (c) on-resonant spectrum of polycrystal, (d) calculated on-resonant spectrum with and without convolution by a Gaussian ($\sigma$=0.5 eV). Dotted lines denote satellite structure. Arrows show energies of Ni 3$d$ $\rightarrow$ 3$p$ fluorescence.}
\end{figure}
 
     Figure 3 shows comparison between on- and off-resonant spectra. The spectra of (110) single crystal are shown in addition to the on-resonant spectrum of polycrystal. The spectra of single crystal show narrower main peak than that of polycrystal, because these were observed in angle resolved mode. In the on-resonance spectra of both samples, two satellite structures are observed at about 2.5 and 4.2 eV as indicated by the dotted lines, while the off-resonant spectrum does not show. A fluorescence component is expected in the on-resonance spectrum at the energy position marked by arrow in Fig. 3, but it is very weak compared with other structures. The spectrum at bottom shows the calculation result \cite{r35} discussed in the following.
 
     We now discuss the origin of the satellite structures. We think the satellite structures are not caused from k-dependence of other components, because Ni 4$sp$ peak is observed broadly in both sample at around 10 eV that is sufficiently higher than the satellite energy. Possibility of direct transition that is observed in UVBIS spectra \cite{r25} can be neglected, because the excitation energy in this study is much higher than UVBIS.

     Since the satellites are observed near absorption edge, it is possible that the structure is caused by a many-body effect, as suggested by Tanaka and Jo.\cite{r35} The spectrum at bottom of Fig. 3 shows RIPES spectra of Ni metal calculated by impurity Anderson model including many-body configuration interaction effect. In the calculation, the initial state of Ni metal consists of 3$d^8$, 3$d^9$ and 3$d^{10}$ configurations, and the IPES spectrum consists of the three structures arising from the bonding, non-bonding and anti-bonding states of the 3$d^9$ and 3$d^{10}$ configurations. The main peak near Fermi edge corresponds to the bonding state and it shows Fano-type resonance, while non-bonding and anti-bonding peaks at 2.5 and 4.2 eV are resonantly enhanced at absorption edge. In this calculation, band effect is not included. If proper band effect is included in this calculation, the non-bonding peak would become wide as observed in experimental results. The intensity changes in this calculation are shown in Fig. 2. The calculated results seem to qualitatively well-describe the intensity change of main peak. From the comparison between the observed and calculated spectra, the weight of 3$d^8$ in Ni metal is estimated to be at least 10 \%.

     As mentioned before, a satellite called the "two-hole-bound state" is observed at 6 eV in resonant PES spectra. The satellite arises from 3$d^8$ dominant states, while the main peak corresponds to 3$d^9$ dominant states. The non-bonding state is not observed in PES spectra. The satellite energy of 6 eV in PES is larger than that of RIPES in this study. This is attributed to the fact that the satellite in PES has 3$d^8$ configuration and Coulomb interaction between two holes is more effective, while the satellite in RIPES has 3$d^9$ configuration. Furthermore, in case of PES spectra that have 3$d^8$ and 3$d^9$ final states, the multiplet splitting of the 3$d^8$ configuration is larger than the hybridization energy, so that the separation of the anti-bonding state from the non-bonding 3$d^8$ state is not obvious. On the other hand, there is no multiplet splitting due to Coulomb interaction in the final states of IPES, because the final states have 3$d^9$ and 3$d^{10}$ configurations. Thus, the non-bonding state would become observable in IPES.

     In conclusion, we could observe RIPES spectra of Ni metal across the Ni 3$p$-3$d$ absorption edge. Satellite structures of Ni 3$d$ band are observed at about 2.5 and 4.2 eV. The excitation spectrum of Ni 3$d$ state shows Fano type resonance across the Ni 3$p$ absorption edge. The results are a direct evidence for existence of 3$d^8$ configuration in the initial state of Ni metal. The satellites are described well by the cluster-model calculation including many-body configuration interaction effects. This result must help for understanding the ferromagnetism on Ni metal.

     The authors thank Prof. Sakisaka, Prof. Kato and Dr. Chainani for useful discussions. This work is partly supported by a Grant-in-Aid for Scientific Research from the Ministry of Education, Science, Sport and Culture.

\end{document}